\documentclass[pdftex, aps, prd, letterpaper, superscriptaddress, nofootinbib, twocolumn,
               preprintnumbers, longbibliography]{revtex4-1}
\pdfoutput=1
\usepackage{amsmath,amssymb,amsbsy}
\usepackage[utf8]{inputenc}
\usepackage{graphicx,stix}
\usepackage{soul,xcolor,topcapt}
\usepackage[colorlinks=true, linkcolor=blue, filecolor=blue, urlcolor=blue, citecolor=blue, pdftex, plainpages=false]{hyperref}
\usepackage[ruled,lined,boxed]{algorithm2e}

\DontPrintSemicolon
\SetKwInOut{Parameter}{Parameters}

\begin{document}
\preprint{FERMILAB-PUB-24-0042-V}
\title{Constrained curve fitting for semi-parametric models with radial basis function networks}

\author{Anna Hasenfratz}
\affiliation{Department of Physics, University of Colorado, Boulder, Colorado 80309, USA}
\author{Curtis T.~Peterson}
\email{curtis.peterson@colorado.edu}
\affiliation{Department of Physics, University of Colorado, Boulder, Colorado 80309, USA}

\begin{abstract}
Common to many analysis pipelines in lattice  gauge theory and the broader scientific discipline is the need to fit  a semi-parametric model to data. We propose a fit method that utilizes a radial basis function network to approximate the non-parametric component of such models. The approximate parametric model is fit to data using the basin hopping global optimization algorithm. Parameter constraints are enforced through Gaussian priors. The viability of our method is tested by examining its use in a finite-size scaling analysis of the $q$-state Potts model and $p$-state clock model with $q=2,3$ and $p=4,\infty$. 
\end{abstract}
\maketitle

\section{Introduction}\label{sec:intro}

Fitting data to a semi-parametric model is necessary in many scientific analysis pipelines.  Well-known examples from lattice gauge theory and condensed matter include spectroscopy and finite-size scaling (FSS). In both cases, the model in question contains a parametric component, from which physically relevant parameters of the model are extracted, in addition to an a priori unknown non-parametric component that is typically estimated from a generic parametric ansatz.  

In a standard FSS curve collapse analysis in the vicinity of a 2nd-order  phase transition, the goal is to determine the critical exponents of some scaling observable $O$, such as a response function. Renormalization group analysis in finite volume predicts at leading order
\begin{equation*}\label{eqn:2nd_order_scaling}
    O(K,N_{s}) \sim N_{s}^{\gamma_{O}}\mathcal{F}_{O}\Big(\big(K/K_c-1\big)N_{s}^{1/\nu}\Big)  \ \ \ \ (K\rightarrow K_{c}),
\end{equation*}
 where $K$ refers to the coupling or temperature that becomes critical at $K_c$, $N_{s}$ is  the linear size of the system\footnote{The lattice volume $V_d=N_{s}^d$ is a dimensionless quantity for symmetric lattices in $d$ dimensions.}, $\nu$ is the critical exponent of the correlation length, and $\gamma_O$ denotes the scaling dimension of the operator $O$. The function $\mathcal{F}_{O}$ is a \textit{universal scaling function}. The functional form of the scaling function $\mathcal{F}_{O}$ is often non-parametric and a priori unknown. The FSS curve collapse analysis attempts to find $K_c$ and the critical exponents by requiring that $O(K,N_{s})\,N_{s}^{-\gamma_{O}}$ is described by a unique function of $(K/K_c-1)\,N_{s}^{1/\nu}$.
It is common to  approximate $\mathcal{F}_{O}$ with some parametric function; e.g., a polynomial or ratio of polynomials.

Another example of fitting to a semi-parametric model emerges in spectroscopy. The ground state amplitude $\tilde{\mathcal{A}}_{0}$ and energy $\tilde{E}_{0}$ in units of the lattice spacing $a$ are extracted from a two-point correlation function $G(\tilde{t})$ using the ansatz 
\begin{equation*}
    G(\tilde{t})=\tilde{\mathcal{A}}_{0}\exp\big(\mbox{-}\tilde{E}_{0}\tilde{t}\big)+\sum_{i=1}^{\infty}\tilde{\mathcal{A}}_{i}\exp\big(\mbox{-}\tilde{E}_{i}.\tilde{t}\big)
\end{equation*}
The infinite sum over the excited-state contributions to $G(\tilde{t})$ is non-parametric in the sense that its exact evaluation requires knowledge of an infinite number of excited-state amplitudes $\tilde{\mathcal{A}}_i$ and energies $\tilde{E}_i$. Estimating $\tilde{\mathcal{A}}_{0}$ and $\tilde{E}_{0}$ requires truncating the excited-state sum, with the order of the truncation chosen such that including higher-order terms minimally impacts the estimate of $\tilde{\mathcal{A}}_0$ and $\tilde{E}_{0}$. 

Other examples of semi-parametric models in lattice field theory, condensed matter physics and the broader scientific domain are abound. Hence, it is desirable to have on hand a class of expressive functions that can faithfully represent the non-parametric component of such models. As universal function approximators,  radial basis function networks (RBFNs) may be just the right tool. For RBFNs to be practically applicable in the high-precision setting of modern lattice gauge theory calculations, one must be able to
\begin{enumerate}
    \item assess quality of fit and model selection criteria,
    \item have a method for estimating correlated statistical uncertainties directly from a single fit, and
    \item accommodate the imposition of statistical constraints and domain-specific knowledge.
\end{enumerate}
 We address these needs using the robust framework of Bayesian statistics and efficient implementation of the basin hopping global optimization algorithm \cite{wales1997global}. 

We test the efficacy of our approach on various finite-size scaling analyses of the $2,3$-state Potts model and $4,\infty$-clock model. Though we focus on FSS for demonstration purposes, we want to stress that both the method and the network architecture that we deploy are broadly applicable to a variety of problems. We emphasize that our goal in this paper is not the precise determination of critical parameters, but to demonstrate the robust applicability of our RBFN-based  method.  We have made our code publicly available to facilitate the deployment and modification of the method \cite{Peterson_SwissFit}. 

The present paper is laid out as follows. In Sec. \ref{sec:fss}, we review the structure of radial basis function networks and the tools of finite-size scaling, introducing our least-squares procedure for fitting RBFNs to data at the end. We review the $q$-state Potts and $p$-state clock models in Sec. \ref{sec:models}. In Sec. \ref{sec:curve_collapse}, we investigate the use of RBFNs in curve collapse analyses of the 2- and 3-state Potts models, along with the 4- and $\infty$-state clock model. We compare RBFN  interpolators against standard polynomial-based interpolations in Sec. \ref{sec:interp_comparison}. We wrap up in Sec. \ref{sec:conclusion} with conclusions and outlook. In Appendix \ref{sec:interpolation}, we briefly explore the use of RBFNs for direct interpolation.

\section{Finite size scaling with radial basis function networks}\label{sec:fss} 

\begin{figure}
    \centering
    \includegraphics[width=\columnwidth]{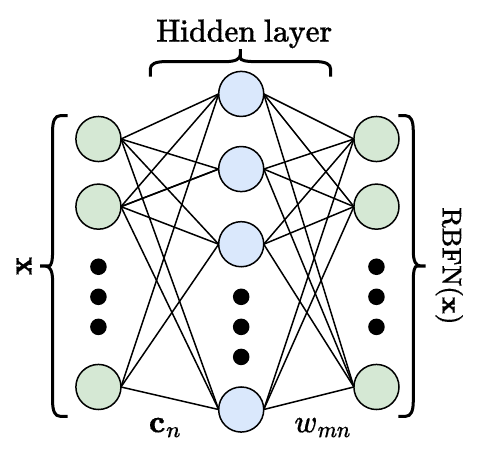}
    \caption{Illustration of a radial basis function network (RBFN). The total number of center parameters $\mathbf{c}_{n}$, counting the components of each $\mathbf{c}_{n}$, is equal to the number of connections between the input nodes (green circles on left; a.k.a., input \textit{features}) and hidden nodes (blue circles). The number of weights is equal to the number of connections between the hidden nodes and output nodes (green circles on right; a.k.a., output \textit{features}).}
    \label{fig:network}
\end{figure}

\subsection{Radial basis function networks}\label{sec:anns}

Radial basis function networks (RBFNs) are artificial neural networks that possess a single hidden layer (see Fig. \ref{fig:network}). The activation function of the hidden layer is a radial basis function (RBF) $\Phi(\cdot)$ and the output of the full network is
\begin{equation}
    \mathrm{RBFN}_{m}(\mathbf{x}) \equiv \sum_{n=1}^{N}w_{mn}\Phi\big(\mbox{-}\beta_{n}^2||\mathbf{x}-\mathbf{c}_{n}||^2\big)+b_{m},
\end{equation}
where $N$ is the number of nodes in the RBFN's hidden layer, $w_{mn}$ are the network \textit{weights}, $b_{m}$ are the network \textit{biases}, $\beta_{n}$ are the RBF \textit{bandwidths}, and $\mathbf{c}_{n}$ are the RBF \textit{centers} \cite{ghosh2001overview}. We denote the parameters of the RBFN as $\Theta_{\mathrm{RBFN}} \equiv \{w_{mn},b_{m},\beta_{n},\mathbf{c}_{n}\}$. There are many choices for the RBFN activation. In this work, we choose the RBF to be exponential
\begin{equation*}
    \Phi(\cdot)=\exp(\cdot);
\end{equation*}
e.g., the activation function has a Gaussian profile. Radial basis function networks are specially designed for function approximation. According to the \textit{universal approximation theorem} for RBFNs, the approximation accuracy of an RBFN scales with the number of nodes in its hidden layer \cite{park1991universal}. As such, RBFNs could be a useful multitool for approximating non-parameteric functions.

\subsection{Finite size scaling with a radial basis function network}

\begin{table*}[tb]
\centering
\begin{tabular}{c@{\extracolsep{1pt}~}cccccccccccc}
\hline\hline
& \multicolumn{3}{c}{$2$-state Potts model} & \multicolumn{3}{c}{$3$-state Potts model} \\
\cline{2-4}\cline{5-7}
 Critical parameter & $U_{4,\mathrm{Potts}}^{(q)}$ & $\chi_{\mathrm{Potts}}^{(q)}$ & Exact & $U_{4,\mathrm{Potts}}^{(q)}$ & $\chi_{\mathrm{Potts}}^{(q)}$ & Exact \\
\cline{0-0}\cline{2-4}\cline{5-7}
$K_{\mathrm{Potts}}^{(q)}$    &  0.881363(15)  &  0.881363(28)  & $\log\big(1+\sqrt{2}\big)$ & 1.00518(15) & 1.005007(48) &  $\log\big(1+\sqrt{3}\big)$ \\
$\nu$    &  0.9995(27)  &  0.9979(40)  & 1 &  0.833(34) & 0.820(23) & 5/6 \\
$\eta$    &  ---  &   0.2496(29) & 1/4 & --- &  0.2713(80) & 4/15 \\
\hline\hline
\end{tabular}
\caption{Comparison of our RBFN-based estimates of $K_{\mathrm{Potts}}^{(q)},\nu$ and $\eta$ critical parameters for $q=2,3$ from a curve collapse analysis of $U_{4,\mathrm{Potts}}^{(q)}$ and $\chi_{\mathrm{Potts}}^{(q)}$. Exact critical parameters are from Ref. \cite{Wu:1982ra}. Predicted critical parameters from Figs. \ref{fig:ising_final_result}-\ref{fig:potts3_final_result}.} \label{tab:potts_parameters}
\end{table*}

\begin{table*}[tb]
\centering
\begin{tabular}{c@{\extracolsep{1pt}~}cccccccccccc}
\hline\hline
& \multicolumn{3}{c}{$4$-state clock model} & \multicolumn{3}{c}{$\infty$-state clock (XY) model} \\
\cline{2-4}\cline{5-7}
 Critical parameter & $U_{4,\mathrm{clock}}^{(p)}$ & $\chi_{\mathrm{clock}}^{(p)}$ & Exact & $U_{4,\mathrm{clock}}^{(p)}$ & $\chi_{\mathrm{clock}}^{(p)}$ & Literature/Exact \\
\cline{0-0}\cline{2-4}\cline{5-7}
$K_{\mathrm{clock}}^{(p)}$    &  0.881379(17)  &  0.881430(66)  & $\log\big(1+\sqrt{2}\big)$  &   1.126(10)  &   1.1160(86)  & 1.1199... \\
$\zeta$    &  ---  &  ---  & --- &  1.6(1.2)  &  1.45(62)  & 1.5... \\
$\nu$    &  0.9976(41)  &  1.001(11)  & 1 &  0.55(20)  &  0.526(92)  & 1/2 \\
$\eta$    &  ---  &   0.2510(39) & 1/4 &  ---  &  0.2513(85) & 1/4 \\
\hline\hline
\end{tabular}
\caption{Comparison of our RBFN-based estimates of $K_{\mathrm{clock}}^{(p)},\nu$, $\eta$, and $\zeta$ critical parameters for $p=4,\infty$ from a curve collapse analysis of $U_{4,\mathrm{clock}}^{(p)}$ and $\chi_{\mathrm{clock}}^{(p)}$. The exact critical parameters are from Refs. \cite{Wu:1982ra, Kosterlitz:1974sm}. Values from the literature for $K_{\mathrm{clock},c}^{(\infty)}$ and $\zeta$ are from Refs. \cite{Kosterlitz:1973xp, hasenbusch2005two, komura2012large, nguyen2021superfluid, Sale:2021xsq}. Predicted critical parameters from Figs. \ref{fig:clock_final_result}-\ref{fig:xy_final_result}.} \label{tab:clock_parameters}
\end{table*}

In the vicinity of a continuous phase transition at the critical value $K_c$ of a macroscopic parameter $K$\footnote{E.g., the temperature $T$ in an equilibrium statistical system or bare gauge coupling $g_{0}^2$ in a gauge-fermion system at zero-temperature.}, finite volume observables $O\big(K, N_{s}\big)$ scale at leading order as
\begin{equation}\label{eqn:scaling}
    O\big(K, N_{s}\big) \sim N_{s}^{\gamma_{O}} \mathcal{F}_{O}(x) \ \ \ \ (K \rightarrow K_c),
\end{equation}
where  $x=x(K,N_{s})$ is a scaling variable. The scaling function $\mathcal{F}_{O}$ is an a priori unknown universal function that depends on $O$. The exponent $\gamma_{O}$ is the anomalous dimension of the operator $O$. If the phase transition is 2nd-order, 
\begin{equation}\label{eqn:1st_2nd}
    x\big(K, N_{s}\big) = \big(K/K_c-1\big)N_{s}^{1/\nu},
\end{equation}
where $\nu$ is the universal critical exponent of the correlation length $\xi(K)$. The same scaling is expected to hold even for first-order phase transitions, with $\nu=1/d$ \cite{Nienhuis:1975zs}. If the phase transition is $\infty$-order, like the Berezinsky-Kosterlitz-Thouless (BKT) phase transition of the 2-dimensional XY model \cite{Kosterlitz:1974sm}, the scaling variable is
\begin{equation}\label{eqn:bkt_scaling}
    x\big(K, N_{s}\big) = N_{s}\exp\Big(\mbox{-}\zeta \big(K/K_c-1\big)^{-\nu}\Big),
\end{equation}
where $\zeta$ is non-universal constant and $\nu$ is a universal critical exponent.

By simulating the system 
in the vicinity of  $K_c$ on multiple volumes $N_{s}^{d}$, it is possible to extract the critical parameters in Eqns. \ref{eqn:scaling}-\ref{eqn:bkt_scaling} by performing a simultaneous fit of $O\big(K,N_{s}\big)$ using several $(K,N_{s})$ values. When the critical parameters are correctly identified, the scaling function $\mathcal{F}_{O}(x)$ becomes independent of the volume, a phenomenon referred to as \textit{curve collapse}. The specific form of $\mathcal{F}_{O}$ is not relevant; it is non-parametric. To estimate the critical parameters from data, we parameterize $\mathcal{F}_{O}$ with an RBFN and determine the parameters of the RBFN $\Theta_{\mathrm{RBFN}}$ as part of the curve collapse. 

Our RBFN-based fits to $\mathcal{F}_{O}$ are performed by minimizing an augmented $\chi^2_{\mathrm{aug.}}$ \cite{Lepage:2001ym, Jay:2020jkz}. We discuss our definition of $\chi^2_{\mathrm{aug.}}$ in Appendix \ref{sec:map}. We partially  control for overfitting by including in $\chi^2_{\mathrm{aug.}}$ a term of the form
\begin{equation}\label{eqn:ridge_prior}
    \chi^2_{\mathrm{ridge}} = \frac{1}{\lambda^2}\sum_{mn}w_{mn}^2,
\end{equation}
which we refer to as a \textit{ridge regression prior}, since it appeared first in the literature on ridge regression \cite{phillips1962technique, tikhonov1963solution, hoerl1970ridge, murphy2023probabilistic}. In the machine learning literature, adding terms of the form of Eqn. \ref{eqn:ridge_prior} to the loss is referred to as \textit{L2-regularization} or \textit{weight decay} \cite{evgeniou2000regularization, burkov2019hundred}. We also add logarithmic constraints in the form of priors to $\chi^2_{\mathrm{aug.}}$ to force positivity on the parameters $K_{c},\nu,\zeta,\gamma_{O}$ in Eqns. \ref{eqn:scaling}-\ref{eqn:bkt_scaling}. We optimize $\chi^2_{\mathrm{aug.}}$ using the basin hopping global optimization algorithm described in Appendix \ref{sec:bhop} \cite{wales1997global}. Additionally, we estimate $\lambda$ using the surrogate-based empirical Bayes procedure described in Appendix \ref{sec:empirical_bayes}. Artificial neural networks with parameters that are estimated from an augmented $\chi^2$ are often referred to as \textit{Bayesian artificial neural networks} \cite{murphy2023probabilistic}.

We note that the use of artificial neural networks for curve collapse was also explored in Ref. \cite{Yoneda:2022bng} using a feedforward neural network. Though we do not illustrate it in this work, we find that feedforward neural networks with Gaussian error linear activation units produce good curve collapse fits. However, we find that it is difficult to perform a stable empirical Bayes analysis using feedfoward neural networks with the present strategy. Nonetheless, our fit software provides support for fitting with feedforward neural networks \cite{Peterson_SwissFit}. The authors of Ref. \cite{Yoneda:2022bng} have also made their code publicly available.

\section{Summary of the investigated Models}\label{sec:models}

We illustrate the efficacy of our RBFN-based fit method by studying the critical properties of several 2-dimensional spin models. This section summarizes the relevant models.

\subsection{The q-state Potts model}

The $q$-state Potts model is a generalization of the Ising model with spin variables taking integer values $s_i\in\{1,...,q\}$ \cite{Potts:1951rk, Wu:1982ra, beffara2012self}. 
The reduced Hamiltonian is defined as
\begin{equation*}
    \mathcal{H}_{\mathrm{Potts}}^{(q)}=-K_{\mathrm{Potts}}^{(q)}\sum_{\langle ij \rangle}\delta(s_i,s_j),
\end{equation*}
where $\langle ij \rangle$ denotes a sum over sites $i$ and nearest-neighbors $j$. The Kronecker delta $\delta(s_i,s_j)=1$ when $s_i=s_j$ and $\delta(s_i,s_j)=0$ otherwise. 
We consider the $q$-state Potts model in $d=2$ dimensions with $q=2,3$. The $q=2$ case is equivalent to the Ising model. For all $q \geq 2$, the $q$-state Potts model exhibits a phase transition at\footnote{Note that the critical coupling for $q=2$ differs from the conventional  Ising model coupling as $K_{\mathrm{Potts},c}^{(2)}=2K_{\mathrm{Ising},c}$} \cite{beffara2012self}
\begin{equation}\label{eqn:potts_crit_temp}
    K_{\mathrm{Potts},c}^{(q)} = 1/\log \big(1+\sqrt{q}\big),
\end{equation}
where the  correlation length $\xi$ in units of the lattice spacing diverges as
\begin{equation}\label{eqn:2nd_xi}
    \xi\big(K_{\mathrm{Potts}}^{(q)}\big) \propto  \big|K_{\mathrm{Potts}}^{(q)}/K_{\mathrm{Potts},c}^{(q)}-1\big|^{-\nu}.
\end{equation}
 The order parameter that distinguishes the phases  is the magnetization 
\begin{equation}\label{eqn:potts_mag}
M\big(K_{\mathrm{Potts}}^{(q)},N_{s}\big) \equiv \frac{1}{N_{s}^2}\sum_{i}\delta(s_i,1)-1/q.
\end{equation} 
 The phase transition is 2nd-order for $q \leq 4$ and 1st-order for $q > 4$ \cite{duminil2016discontinuity, duminil2017continuity}. We list the critical exponents $\nu,\eta$\footnote{$\eta$ is the critical exponent of the  wave function, related to the critical exponent of the magnetic susceptibility as $\gamma/\nu = 2-\eta$.} and the critical couplings $K_{\mathrm{Potts}}^{(q)}$ for the  2- and 3-state systems in Table \ref{tab:potts_parameters}. We simulate the $q=2,3$ system around $K_{\mathrm{Potts},c}^{(q)}$ using the Wolff cluster algorithm provided by the \texttt{Julia}-based \texttt{SpinMonteCarlo} library \cite{Wolff:1988uh, Yuichi:2019}

\begin{figure}
    \centering
    \includegraphics[width=\columnwidth]{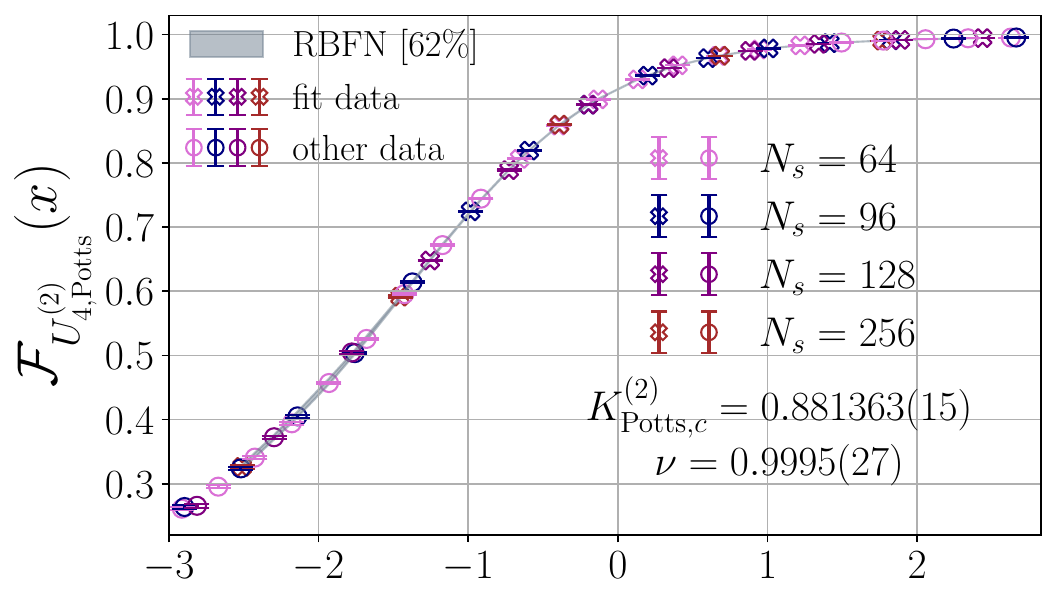}
    \includegraphics[width=\columnwidth]{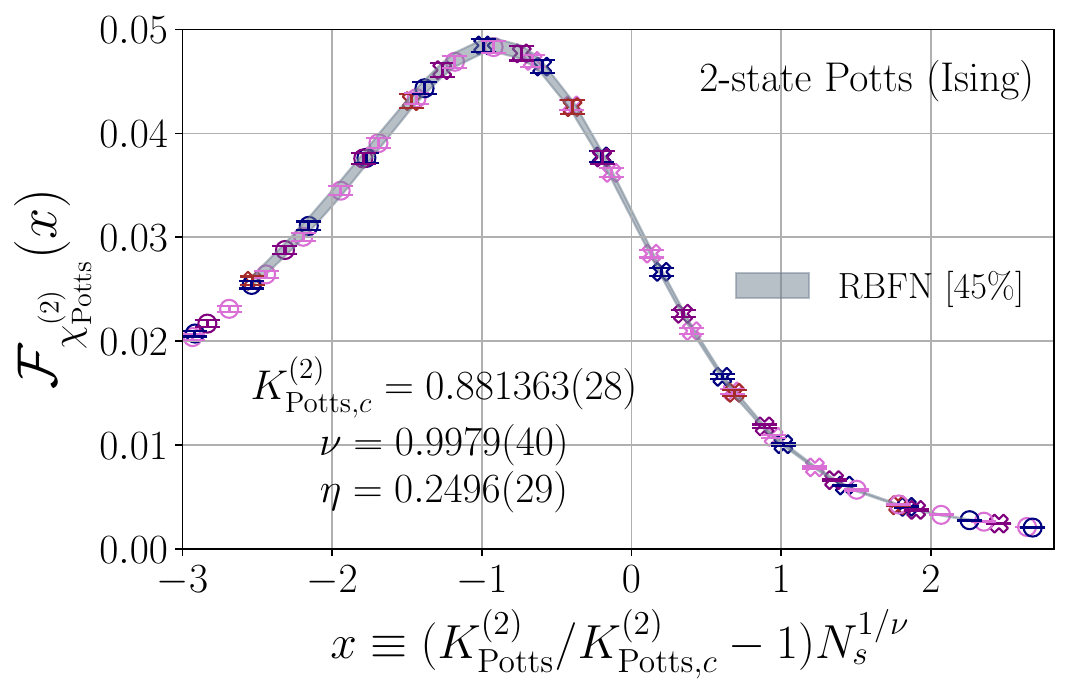}
    \caption{RBFN-based curve collapse analysis of the 2-state Potts (Ising) model using the Binder cumulant $U_{4,\mathrm{Potts}}^{(2)}$ (top panel) and the magnetic susceptibility $\chi_{\mathrm{Potts}}^{(2)}$ (bottom panel). The curve collapse uses $N_{s}=64$ (pink), 96 (blue), 128 (purple), and 256 (red) volumes in the coupling range  $0.87 \leq K_{\mathrm{Potts}}^{(2)} \leq 0.90$. Data used in the curve collapse are marked with an open $\times$ (fit data); otherwise, they are marked with an open $\circ$ (other data). The scaling function $\mathcal{F}_{O}$ predicted by the RBFN is plotted as a grey band. The width of the band corresponds to the predicted error. The RBFN in the top panel has two nodes in its hidden layer and the RBFN in the bottom panel has three.}
    \label{fig:ising_final_result}
\end{figure}

\begin{figure}
    \centering
    \includegraphics[width=\columnwidth]{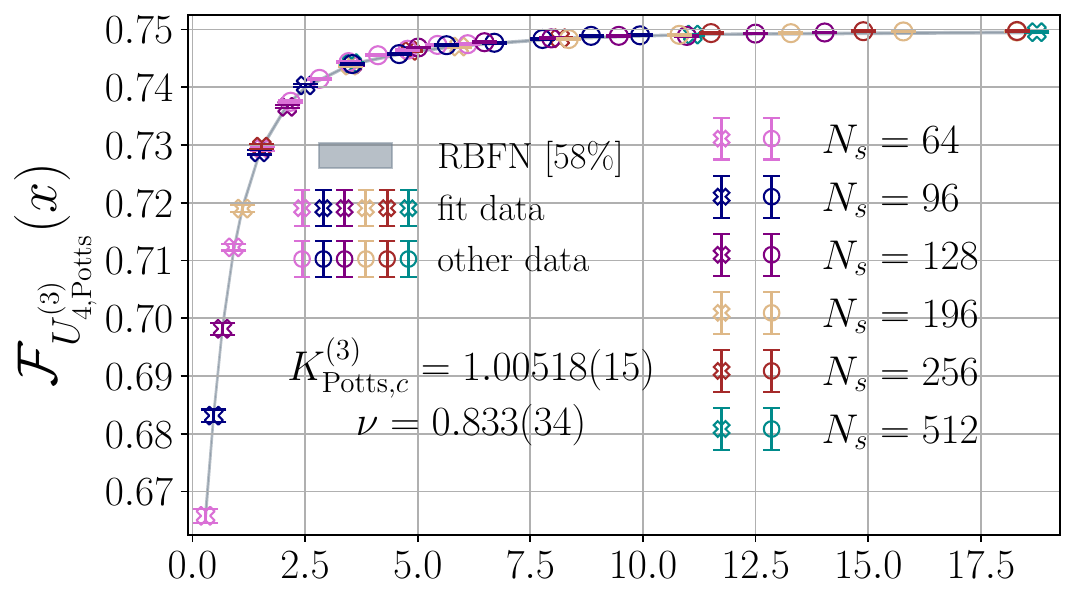}
    \includegraphics[width=\columnwidth]{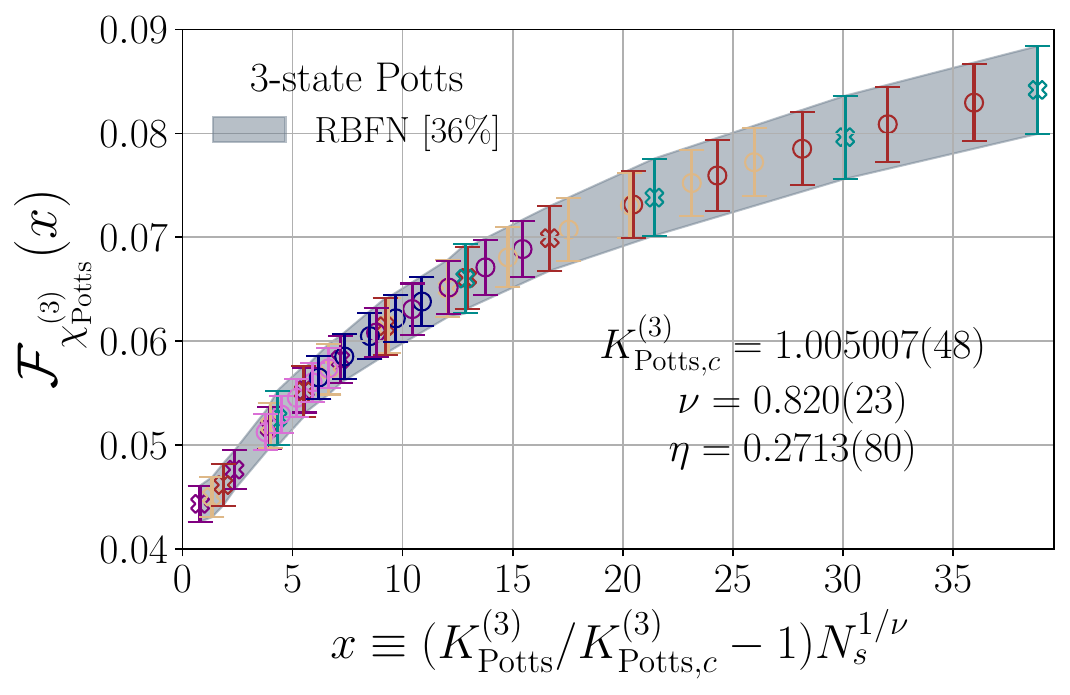}
    \caption{RBFN-based curve collapse analysis of the 3-state Potts model using the Binder cumulant $U_{4,\mathrm{Potts}}^{(3)}$ (top panel) and order parameter susceptibility $\chi_{\mathrm{Potts}}^{(3)}$ (bottom panel). The curve collapse uses $N_{s}=64$ (pink), 96 (blue), 128 (purple), 196 (tan), 256 (red), and 512 (cyan) volumes for $U_{4,\mathrm{Potts}}^{(3)}$ and $N_{s}=128, 196, 256$, and 512 volumes for $\chi_{\mathrm{Potts}}^{(3)}$ (same scheme as $U_{4,\mathrm{Potts}}^{(3)}$). The $K_{\mathrm{Potts}}^{(3)}$ values used in both curve collapse analyses are in the range $1.005 \leq K_{\mathrm{Potts}}^{(3)} \leq 1.018$ for $U_{4,\mathrm{Potts}}^{(3)}$ and $1.005 \leq K_{\mathrm{Potts}}^{(3)} \leq 1.026$ for $\chi_{\mathrm{Potts}}^{(3)}$. Data used in the curve collapse are marked with an open $\times$ (fit data); otherwise, they are marked with an open $\circ$ (other data). The scaling function $\mathcal{F}_{O}$ predicted by the RBFN is plotted as a grey band. The width of the band corresponds to the predicted error. The RBFN in both panels has two nodes in its hidden layer.}\label{fig:potts3_final_result}
\end{figure}

\subsection{The p-state clock model}
The $p$-state clock model is 
a discrete version of the XY model. The spin variables are angles  $\theta_{i}=2\pi n_i/p$ for $1 \leq n_i \leq p$ and the reduced Hamiltonian is defined as 
\begin{equation*}
    \mathcal{H}_{\mathrm{clock}}^{(p)}=-K_{\mathrm{clock}}^{(p)}\sum_{\langle ij \rangle}\cos\big(\theta_{i}-\theta_{j}\big).
\end{equation*}
The $p=2$ case is the Ising model and the $p\rightarrow\infty$ limit is equivalent to the XY-model.  We consider the $p$-state clock model in $d=2$ dimensions with $p=4,\infty$. The 4-state clock model is in the Ising universality class \cite{Elitzur:1979uv}. It has a 2nd-order phase transition at \cite{Ortiz:2012zz}
\begin{equation}\label{eqn:4_state_crit_param}
    K_{\mathrm{clock}}^{(4)}=1/\log\Big(1+\sqrt{2}\Big),
\end{equation}
where the correlation length in units of the lattice spacing diverges as Eqn. \ref{eqn:2nd_xi} with the replacement $K_{\mathrm{Potts}}^{(q)}\rightarrow K_{\mathrm{clock}}^{(4)}$.  The  $\infty$-state clock (XY) model has a topology-driven $\infty$-order Berezinsky-Kosterlitz-Thouless (BKT) phase transition at $K_{\mathrm{clock},c}^{(\infty)} \approx 1.1199$ \cite{Kosterlitz:1973xp, Kosterlitz:1974sm, hasenbusch2005two, komura2012large, nguyen2021superfluid, Sale:2021xsq}, where the  correlation length in units of the lattice spacing diverges as
\begin{equation}\label{eqn:bkt_xi}
    \xi\big(K_{\mathrm{clock}}^{(\infty)}\big) \propto  \exp\Big(\zeta \big|K_{\mathrm{clock}}^{(\infty)}/K_{\mathrm{clock},c}^{(\infty)}-1\big|^{-\nu}\Big).
\end{equation}
The known critical parameters $\zeta,\nu$ and $\eta$ for the $4$- and $\infty$-state clock model are listed in Table \ref{tab:clock_parameters}, along with the critical couplings $K_{\mathrm{clock}}^{(p)}$. We simulate the $4$-state clock model using the \texttt{SpinMonteCarlo} library's implementation of the Wolff cluster algorithm. We simulate the $\infty$-state clock (XY) model using \texttt{Nim}-based \texttt{QuantumEXpressions} library's implementation of the heat bath algorithm \cite{Wolff:1988uh, Osborn:2017aci, Yuichi:2019}.

\section{Curve collapse results}\label{sec:curve_collapse}

\begin{figure}
    \centering
    \includegraphics[width=\columnwidth]{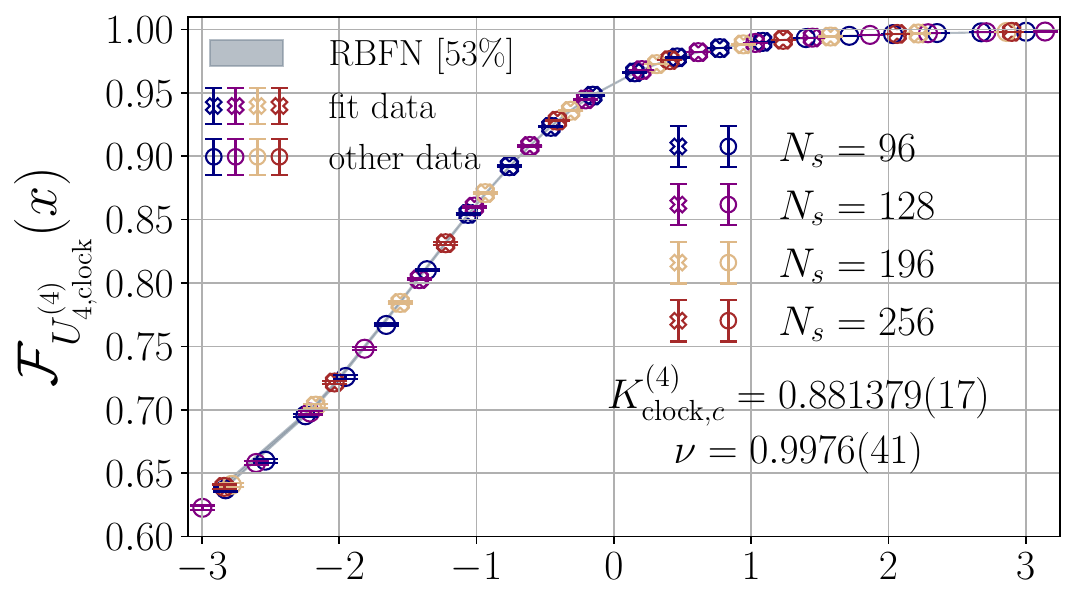}
    \includegraphics[width=\columnwidth]{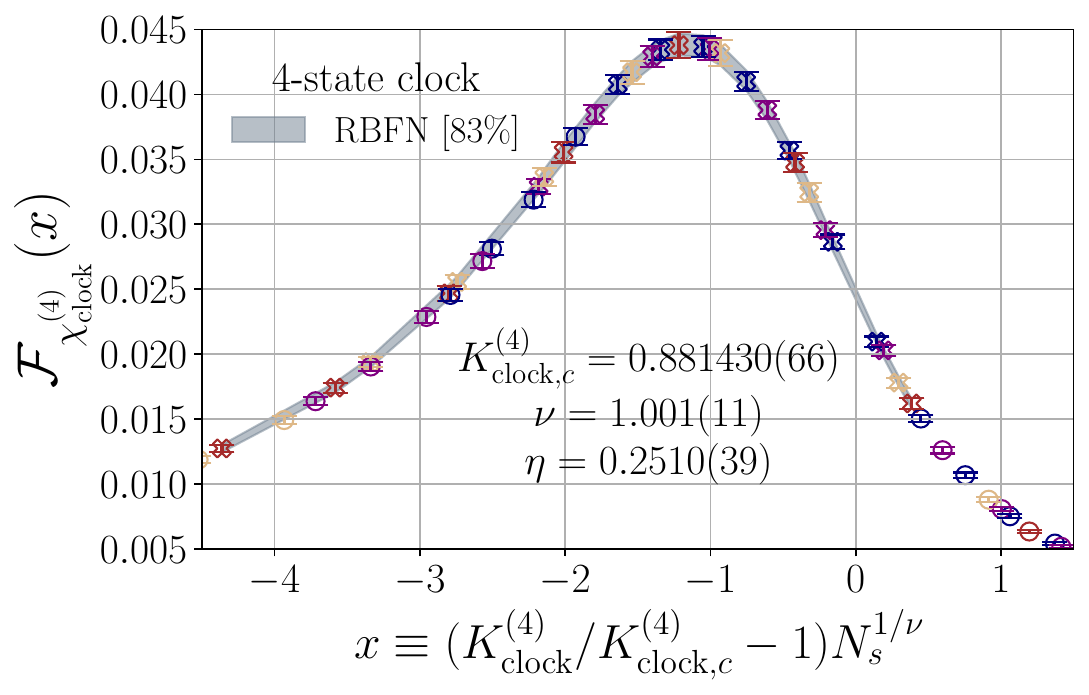}
    \caption{RBFN-based curve collapse of the $4$-state clock model using the Binder cumulant $U_{4,\mathrm{clock}}^{(4)}$ (top panel) and connected magnetic susceptibility $\chi_{\mathrm{clock}}^{(4)}$ (bottom panel). The curve collapse uses $N_{s}=96$ (blue), 128 (purple), 196 (tan), 256 (red). The $K_{\mathrm{clock}}^{(4)}$ values used in the curve collapse of $U_{4,\mathrm{clock}}^{(4)}$ are in the range $0.870 \leq K_{\mathrm{clock}}^{(4)} \leq 0.893$ and the $K_{\mathrm{clock}}^{(4)}$ values used for $\chi_{\mathrm{clock}}^{(4)}$ are in the range $0.870 \leq K_{\mathrm{clock}}^{(4)} \leq 0.885$.  Data used in the curve collapse is marked with an open $\times$ (fit data); otherwise, it is marked with an open $\circ$ (other data). The scaling function $\mathcal{F}_{O}$ predicted by the RBFN is plotted as a grey band. The width of the band corresponds to the predicted error. The RBFN in both panels has three nodes in its hidden layer.}\label{fig:clock_final_result}
\end{figure}

\begin{figure}
    \centering
    \includegraphics[width=\columnwidth]{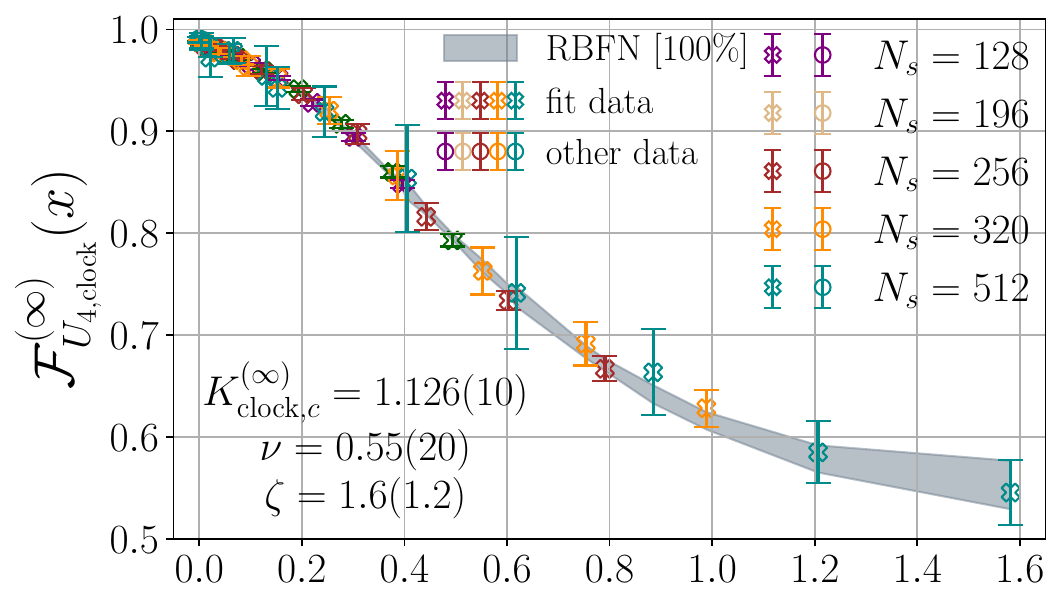}
    \includegraphics[width=\columnwidth]{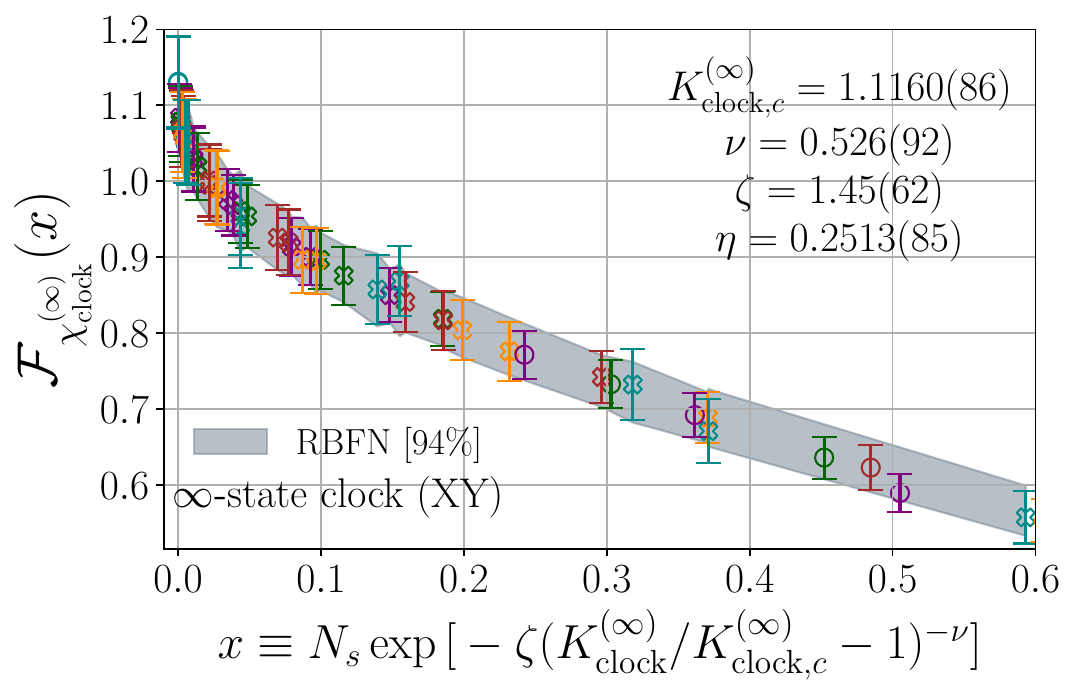}
    \caption{RBFN-based curve collapse analysis of the $\infty$-state clock (XY) model using the Binder cumulant $U_{4,\mathrm{clock}}^{(\infty)}$ (top panel) and connected magnetic susceptibility $\chi_{\mathrm{clock}}^{(\infty)}$ (bottom panel). The curve collapse uses $N_{s}=128$ (purple), $196$ (tan), 256 (red), 320 (yellow), and 512 (cyan). The $K_{\mathrm{clock}}^{(\infty)}$ values used for both curve collapse analyses between 1.005/1.0 for $U_{4,\mathrm{clock}}^{(\infty)}/\chi_{\mathrm{clock}}^{(\infty)}$ and 1.1, 1.1, 1.102, 1.102, and 1.105 for $N_{s}=128,320,256,160$ and $512$, respectively for both observables. Data used in the curve collapse is marked with an open $\times$ (fit data); otherwise, it is marked with an open $\circ$ (other data). The scaling function $\mathcal{F}_{O}$ predicted by the RBFN is plotted as a grey band. The width of the band corresponds to the predicted error. The RBFN in both panels has two nodes in its hidden layer.}
    \label{fig:xy_final_result}
\end{figure}

\subsection{$q$-state Potts model}

We determine the critical parameters $K_{\mathrm{Potts},c}^{(q)}$, $\nu$, and $\eta$ for the 2- and 3-state Potts model from a curve collapse analysis of the Binder cumulant 
\begin{equation}\label{eqn:binder_cumulant}
U_{4,\mathrm{Potts}}^{(q)}\big(K_{\mathrm{Potts}}^{(q)},N_{s}\big)=\frac{1}{2}\Bigg[3-\frac{\langle M^{4} \rangle}{\langle M^{2} \rangle^2}\Bigg]
\end{equation}
and the connected magnetic susceptibility
\begin{equation}\label{eqn:sus}
\chi_{\mathrm{Potts}}^{(q)}\big(K_{\mathrm{Potts}}^{(q)},N_{s}\big)=K_{\mathrm{Potts}}^{(q)}N_{s}^2\big\langle (|M|-\langle |M| \rangle)^2 \big\rangle,
\end{equation} 
where $M\big(K_{\mathrm{Potts}}^{(q)},N_{s}\big)$ is defined in Eqn. \ref{eqn:potts_mag}. In Fig. \ref{fig:ising_final_result}, we show the result of our RBFN-based curve collapse for the $2$-state Potts model using the Binder cumulant (top panel) and connected magnetic susceptibility (bottom panel). We show the same information for the $3$-state Potts model in Fig. \ref{fig:potts3_final_result}.  
In Tab. \ref{tab:potts_parameters}, we compare our prediction for the critical parameters $K_{\mathrm{Potts}}^{(q)},\nu,$ and $\gamma_{\chi_{\mathrm{Potts}}^{(q)}}\equiv 2-\eta$ with their exact values from Ref. \cite{Wu:1982ra}.
Despite the relatively small number of nodes in each hidden layer (only 2-3), the RBFN fits the data well, with p-values in the range $36\%-62\%$ and predictions for the critical parameters that  are in statistical agreement with their exact values from Ref. \cite{Wu:1982ra}.

\begin{table*}[tb]
\centering
\begin{tabular}{c@{\extracolsep{1pt}~}ccccccccccccccc}
\hline\hline
& \multicolumn{2}{c}{$U_{4,\mathrm{Potts}}^{(2)}$} & \multicolumn{2}{c}{$\chi_{\mathrm{Potts}}^{(2)}$} & \multicolumn{2}{c}{$U_{4,\mathrm{clock}}^{(\infty)}$} & \multicolumn{2}{c}{$\chi_{\mathrm{clock}}^{(\infty)}$} \\
\cline{2-3}\cline{4-5}\cline{6-7}\cline{8-9}
 fit result & polynomial & RBFN & polynomial & RBFN & polynomial & RBFN & polynomial & RBFN  \\
\cline{0-0}\cline{2-3}\cline{4-5}\cline{6-7}\cline{8-9}
AIC/($D+k$) & 1.74 & 0.99 & 1.29 & 1.09 & 1.12 & 0.45 & 0.84 & 0.85 \\
$\chi_{\mathrm{aug.}}^2/\mathrm{d.o.f.}$ & 1.89 & 0.89 & 1.32 & 1.00 & 1.16 & 0.27 &  0.74 & 0.68 \\
p-value & 0\% & 62\% & 12\% & 46\% & 17\% & 100\% & 90\% & 94\% \\
\hline
$K_{c}$ & 0.881335(14) & 0.881363(15) & 0.881420(57) &  0.881363(28) & 1.1568(87) & 1.126(10) & 1.1100(63) & 1.1160(86) \\
$\zeta$ & --- & --- & --- & --- & 2.33(24) & 1.6(1.2) & 0.58(19) & 1.45(62) \\
$\nu$   & 0.9954(25) & 0.9995(27) &  0.9913(62) & 0.9979(40) & 0.576(51) & 0.55(20) & 0.683(97) & 0.526(92) \\
$\eta$ & --- & --- & 0.2544(54) & 0.2496(29) & --- & --- & 0.2584(56) & 0.2513(85) \\
\hline\hline
\end{tabular}
\caption{Comparison of curve collapse analysis using a polynomial (Eqn. \ref{eqn:polynomial}) against our RBFN-based ansatz. Curve collapse analysis performed with the Binder cumulant and connected susceptibility of the 2-state Potts model and $\infty$-state clock model.}\label{tab:curve_collapse}
\end{table*}

\subsection{$p$-state  clock model}

We determine the critical parameters $K_{\mathrm{clock},c}^{(p)}$, $\nu$, and $\eta$ for the 4- and $\infty$-state clock model from a curve collapse analysis of the Binder cumulant and magnetic susceptibility. For both models, the Binder cumulant is defined similarly to Eqn. \ref{eqn:binder_cumulant} using the magnitude of the  magnetization vector
\begin{equation}\label{eqn:vec_mag}
    \mathbf{M}\big(K_{\mathrm{clock}}^{(p)},N_{s}\big) \equiv \sum_{i}\big(\cos(\theta_i),\sin(\theta_i)\big)
\end{equation}
in place of the magnetization defined in Eqn. \ref{eqn:potts_mag} for the Potts model. 
We calculate the magnetic susceptibility as in Eqn. \ref{eqn:sus} using the magnitude of the magnetization defined in Eqn. \ref{eqn:vec_mag}. For the $\infty$-state clock model, we obtain a better curve collapse using the  estimator for the  magnetic susceptibility
\begin{equation}\label{eqn:xy_sus}  \chi_{\mathrm{clock}}^{(\infty)}\big(K_{\mathrm{clock}}^{(\infty)},N_{s}\big)=\Big\langle \big|\mathbf{M}\big(K_{\mathrm{clock}}^{(\infty)},N_{s}\big)\big|^2 \Big\rangle
\end{equation}
suggested in Refs. \cite{Gupta:1992zz, ota1992microcanonical}. In Tab. \ref{tab:clock_parameters}, we compare our estimates for the critical parameters of both models, including $\zeta$ for the $\infty$-state clock model, against both exact values and estimates from the literature.

In Figs. \ref{fig:clock_final_result} and \ref{fig:xy_final_result} we show the result of our curve collapse analysis of the Binder cumulant (top panels) and magnetic susceptibility (bottom panels) for the 4- and $\infty$-state clock models, respectively. The p-values of our curve collapse for the 4-state clock model of both observables are $53\%$ and 83\%. For the $\infty$-state clock model, they are 100\% and 94\%, indicating that either the RBFN overfits or the statistical errors of the data entering our curve collapse are overestimated. For both models, both exact results and values from the literature are within $1\sigma$ of our predictions. Notably, our predictions for the $4$-state clock model confirm that it is not only in the Ising universality class but that it also has the same critical coupling as the $2$-state Potts (Ising) model \cite{Elitzur:1979uv, Ortiz:2012zz}.

\section{Comparison against polynomial ansatz}\label{sec:interp_comparison}

In Sec. \ref{sec:curve_collapse}, we have illustrated the robust applicability of our RBFN  fit method. In this section, we contrast the RBFN  with a polynomial ansatz
\begin{equation}\label{eqn:polynomial}
    \mathcal{P}(x)=\sum_{i=0}^{P-1}p_{i}x^{i}
\end{equation}
and ridge regression prior
\begin{equation}\label{eqn:poly_ridge_prior}
    \chi^2_{\mathrm{poly. \ ridge}} = \frac{1}{\lambda^2}\sum_{i=0}^{P-1}p_{i}^2.
\end{equation}
As we do for the RBFN, we calculate $\lambda$ using the surrogate-based empirical Bayes method described in Appendix \ref{sec:empirical_bayes}. Our comparison utilizes the $\chi^2_{\mathrm{aug.}}/\mathrm{d.o.f.}$ and p-value as measures of the quality of fit, along with the Akaike information criterion (AIC) \cite{Jay:2020jkz}
\begin{equation}
    \mathrm{AIC} \equiv \chi^2_{\mathrm{data}}+2k,
\end{equation}
where $\chi^2_{\mathrm{data}}$ is the $\chi^2$ of the data defined in Appendix \ref{sec:map} and $k$ is the total number of fit parameters. A good fit has $\chi^2_{\mathrm{data}} \approx D-k$, where $D$ is the size of the dataset. This translates to $\mathrm{AIC} \approx D+k$. We also compare the accuracy of predictions of critical parameters to their exact values and the literature. In all cases, the total number of parameters in Eqn. \ref{eqn:polynomial} is equal to the total number of parameters of the RBFN that they are being compared against.

In Table \ref{tab:curve_collapse}, we compare our RBFN-based ansatz for the scaling function $\mathcal{F}_{O}$ of the Binder cumulant and magnetic susceptibility for the 2-state Potts (Ising) model and $\infty$-state clock (XY) model against the polynomial ansatz of Eqn. \ref{eqn:polynomial}. The RBFN fits the 2-state Potts data the best, with a $\chi^2_{\mathrm{aug.}}/\mathrm{d.o.f.}$ closest to unity for both observables, along with an AIC/$(D+k)$ that is closest to unity. For the Binder cumulant in the 2-state Potts model, the polynomial ansatz predicts a value for $K_{\mathrm{Potts},c}^{(2)}$ and $\nu$ that is greater than $1\sigma$ away from the exact value of both critical parameters. The same is true for the prediction of $\nu$ from the connected susceptibility in the 2-state Potts model. For the $\infty$-state clock model, both fit ansatz overfit the data, indicating that the errors in the data are possibly overestimated. The predictions for $K_{\mathrm{clock},c}^{(\infty)}, \zeta, \nu$ from the polynomial ansatz are greater than $1\sigma$ away from their estimates in the literature for both observables and the prediction for $\eta$ from the connected susceptibility is greater than $1\sigma$ away from its exact value. All other predictions from both fit ansatz produce results that are within $1\sigma$ of either the literature or their exact values. It is clear that, with our choice of priors on the parameters of both models, the RBFN produces the most consistently correct results for the Binder cumulant and connected susceptibility of the 2-state Potts model and $\infty$-state clock model.

\section{Discussion and conclusion}\label{sec:conclusion}

We have investigated the use of radial basis function networks as a tool for describing the non-parametric component of semi-parameter models. For illustration, we have used an RBFN for curve-collapse analysis of the $2,3$-state Potts model and $4,\infty$-state clock model in Sec. \ref{sec:curve_collapse}. We find that the RBFN fits a variety of disparate curves very well. Most importantly, our RBFN-based fits produce predictions for critical parameters that are consistent with their exact results and the literature. The RBFN tends to perform better than the polynomial-based fit ansatz for the scaling function with ridge regression priors given by Eqn. \ref{eqn:poly_ridge_prior}. 

Though our RBFN-based fit procedure is readily available for direct use in other scientific analyses, several improvements could be made for future applications. These include improvements to the basin hopping global optimization algorithm outlined in Appendix \ref{sec:bhop} and a more complicated, yet efficient, empirical Bayes procedure for constraining more than just the weights of the RBFN. We have made our fit software publicly available for ease of deploying the method presented in this work and future experimentation of both RBFNs and feedforward neural networks.

\begin{acknowledgements}
 Both authors acknowledge support by DOE Grant No.~DE-SC0010005. This material is based upon work supported by the National Science Foundation Graduate Research Fellowship Program under Grant No.~DGE 2040434. The research reported in this work made use of computing and long-term storage facilities of the USQCD Collaboration, which are funded by the Office of Science of the U.S. Department of Energy. We benefited from many comments and discussions during ``The International Symposium on Lattice Field Theory'' at Fermilab, Batavia, Illinois, USA, July 31 - Aug. 04, 2023.
\end{acknowledgements}

\bibliography{./BSM}

\appendix

\section{Augmented $\chi^2$ and parameter estimation}\label{sec:map}
In this work, we model the scaling function $\mathcal{F}_{O}$ as
\begin{equation*}
    \mathcal{F}_{O}(K,N_{s}) \approx N_{s}^{-\gamma_{O}}\mathrm{RBFN}\big(x(K,N_{s})\big)
\end{equation*}
for curve collapse (Sec. \ref{sec:curve_collapse}) or observables $O(K,N_{s})$ at fixed $N_{s}$ as
\begin{equation}
    O(K,N_{s})\approx\mathrm{RBFN}(K,N_{s}),
\end{equation}
for curve fitting/interpolation (Appendix \ref{sec:interpolation}). Let us denote either model as $\mathcal{M}_{\Theta}(\mathbf{x})$ with model parameters $\Theta$. When estimating the scaling function $\mathcal{F}_{O}$, $\Theta$ contains both the critical parameters and the parameters of the RBFN. When directly interpolating observables $O(K,N_{s})$ at fixed $N_{s}$, $\Theta$ contains just the parameters of the RBFN. We estimate $\Theta$ by fitting $\mathcal{M}_{\Theta}$ to a dataset with $D$ inputs $\mathbf{x}^{i}$ and $D$ Gaussian distributed outputs $\mathcal{D}_i \equiv \mathcal{D}(\mathbf{x}^{i})$ ($1\leq i \leq D$). Denoting the covariance of $\{\mathcal{D}_i\}_{1\leq i\leq D}$ as $\Sigma_{\mathcal{D}}$ and the mean of each $\mathcal{D}_{i}$ as $\overline{\mathcal{D}}_{i}$, we define the $\chi^2$ of the data as
\begin{equation}\label{eqn:likelihood}
    \chi^2_{\mathrm{data}}(\Theta) \equiv \sum_{i,j=1}^{D}\big(\mathcal{M}_{\Theta}(\mathbf{x}^{i})-\overline{\mathcal{D}}_i\big)\big[\Sigma_{\mathcal{D}}^{-1}\big]_{ij}\big(\mathcal{M}_{\Theta}(\mathbf{x}^{j})-\overline{\mathcal{D}}_j\big).
\end{equation}
Note that $\chi^2_{\mathrm{data}}$ is proportional to the log \textit{likelihood} of a multivariate Gaussian model. We represent anything that we know about $\Theta$ before fitting $\mathcal{M}_{\Theta}$ to $\{\mathcal{D}_i\}_{1\leq i\leq D}$ by Gaussian priors. The priors  take the form of $C$ constraints $\mathcal{C}^{k}(\Theta)$ with mean $\overline{\mathcal{C}}_{k}$ and covariance $\Sigma_{\mathcal{C}}$ ($1 \leq k \leq C$). We define the $\chi^2$ of the prior as
\begin{equation}\label{eqn:prior}
    \chi^2_{\mathrm{prior}}(\Theta) \equiv \sum_{k,l=1}^{C}\big(\mathcal{C}^{k}(\Theta)-\overline{\mathcal{C}}_{k}\big)\big[\Sigma_{\mathcal{C}}^{-1}\big]_{kl}\big(\mathcal{C}^{l}(\Theta)-\overline{\mathcal{C}}_{l}\big),
\end{equation}
which is proportional to the log of a Gaussian \textit{prior} distribution. According to Bayes' theorem, the log of the \textit{posterior} distribution that is derived from our choice of likelihood (Eqn. \ref{eqn:likelihood}) and prior (Eqn. \ref{eqn:prior}) is proportional to \cite{Lepage:2001ym, Jay:2020jkz, neil2022improved}
\begin{equation}\label{eqn:loss}
    \chi^2_{\mathrm{aug.}}(\Theta) \equiv \chi^2_{\mathrm{data}}(\Theta) + \chi^2_{\mathrm{prior}}(\Theta).
\end{equation}
When we fit $\mathcal{M}_{\Theta}$ to $\{\mathcal{D}_i\}_{1\leq i\leq D}$ with $C$ prior constraints $\mathcal{C}^{k}(\Theta)$, we are calculating a \textit{maximum a posteriori} estimate of $\Theta$, defined by
\begin{equation}
    \Theta^{*} \equiv \mathrm{argmin}_{\Theta} \ \chi^2_{\mathrm{aug.}}(\Theta).
\end{equation}
The MAP estimate $\Theta^{*}$ is also known as the \textit{posterior mode}. Laplace approximation of the posterior about $\Theta^{*}$ yields a convenient approximation of the posterior covariance $\Sigma_{\Theta}$:
\begin{align}\label{eqn:cov_of_params}
    \big[\Sigma_{\Theta}^{-1}\big]_{mn} \approx \sum_{i,j=1}^{D}\frac{\partial \mathcal{M}_{\Theta}(\mathbf{x}_i)}{\partial\Theta_{m}}\big[\Sigma_{\mathcal{D}}^{-1}\big]_{ij}\frac{\partial \mathcal{M}_{\Theta}(\mathbf{x}_j)}{\partial\Theta_{n}}\Bigg|_{\Theta=\Theta^{*}} \notag \\ 
    + \sum_{k,l=1}^{C}\frac{\partial \mathcal{C}^{k}(\Theta)}{\partial\Theta_{m}}\big[\Sigma_{\mathcal{C}}^{-1}\big]_{kl}\frac{\partial \mathcal{C}^{l}(\Theta)}{\partial\Theta_{n}}\Bigg|_{\Theta=\Theta^{*}} 
\end{align}
where $\Theta_{m},\Theta_{n}\in\Theta$. The error in $\Theta$ that is calculated from Eqn. \ref{eqn:cov_of_params} is propagated into $\mathcal{M}_{\Theta}(\mathbf{x})$ using the automatic Gaussian error propagation tools provided by the \texttt{GVar} library \cite{gvar}. Specifically, the error $\sigma^2_{\mathcal{M}_{\Theta}}(\mathbf{x})$ of $\mathcal{M}_{\Theta}(\mathbf{x})$ from the error in the parameters is given by
\begin{equation}\label{eqn:error_in_M}
    \sigma^2_{\mathcal{M}_{\Theta}}(\mathbf{x})=\sum_{m,n=1}^{k}\frac{\partial \mathcal{M}_{\Theta}(\mathbf{x})}{\partial \Theta_{m}}\big[\Sigma_{\Theta}\big]_{mn}\frac{\partial \mathcal{M}_{\Theta}(\mathbf{x})}{\partial \Theta_{m}},
\end{equation}
where $k$ is the total number of model parameters. Correlations between $\mathcal{M}_{\Theta}(\mathbf{x})$ and $\mathcal{M}_{\Theta}(\mathbf{y})$ for any $\mathbf{x},\mathbf{y}$ are calculated and kept track of in an automated manner using \texttt{GVar} \cite{gvar}.


\section{Optimization with basin hopping}\label{sec:bhop}
\begin{algorithm}[t!]
\caption{The basin hopping global optimization algorithm, as implemented in Ref. \cite{wales1997global}. The \texttt{LocalOptimization} step utilizes the trust region reflective local optimization algorithm \cite{branch1999subspace}. We use the \texttt{SciPy} library's implementation of both optimization algorithms \cite{2020SciPy-NMeth}.}\label{alg:cap}
\KwIn{$\Theta_{0},\alpha,T$}
$\Theta \gets \mathrm{\texttt{LocalOptimization}}(\Theta_{0})$\;
$\Theta_{\mathrm{best}} \gets \Theta$\;
\SetInd{0.1em}{2em}
\While{$\Theta_{\mathrm{best}}$ not converged}{
    $\Theta' \gets \mathrm{\texttt{RandomPerturbation}}(\Theta,\alpha)$\;
    $\Theta' \gets \mathrm{\texttt{LocalOptimization}}(\Theta')$\;
    $\Theta \gets \mathrm{\texttt{MetropolisCriterion}}(\Theta,\Theta',T)$ \;
    \If{new $\Theta_{\mathrm{best}}$}{
        $\Theta_{\mathrm{best}} \gets \Theta$
    }
}
\KwOut{$\Theta_{\mathrm{best}}$}
\end{algorithm}

The landscape of $\chi^2_{\mathrm{aug.}}(\Theta)$ in $\Theta$ is complicated. Depending on the problem, the $\chi^2_{\mathrm{aug.}}$ landscape may possess many local optima with their own basins of attraction, along with sharp barriers that separate regions in $\Theta$ space. If one uses a local optimization algorithm to calculate the posterior mode $\Theta^{*}$, it is important to be careful with the initialization $\Theta$; otherwise, the algorithm is bound to converge to one of many local optima. One way out is to utilize a variant of stochastic gradient descent, such as the Adam optimization algorithm or its Nesterov-accelerated counterpart \cite{kingma2017adam, dozat2016incorporating}. However, such algorithms are only efficient on large datasets. Another approach is to utilize a global optimization algorithm. In this work, we deploy the basin hopping (BH) global optimization algorithm first championed in Ref. \cite{wales1997global}. Though we do not illustrate it in this work, we find that basin hopping vastly outperforms other common global optimization algorithms, such as generalized simulated annealing and various metaheuristic evolutionary algorithms \cite{xiang2013generalized, storn1997differential}.

The most basic implementation of BH optimizes $\chi^2_{\mathrm{aug.}}$ by repeatedly performing  a random perturbation (``hop'') of $\Theta\rightarrow\Theta'$ with step size $\alpha$ that is followed by a local optimization of $\chi^2_{\mathrm{aug.}}(\Theta')$ (see Algorithm \ref{alg:cap}). The optimized $\Theta'$ at each step of the algorithm is accepted with probability 
\begin{equation}\label{eqn:acc_prob}
\mathrm{acc. \ prob.}=\exp\Big[\mbox{-}\mathrm{max}\Big(0,\chi^2_{\mathrm{aug.}}(\Theta')-\chi^2_{\mathrm{aug.}}(\Theta)\Big)\Big]^{1/T}
\end{equation} 
at the \texttt{MetropolisCriterion} step of Algorithm \ref{alg:cap}, where $T$ is a hyperparameter that is referred to as the ``temperature''. Our implementation of BH is built on top of the \texttt{SciPy} library's implementation of BH \cite{2020SciPy-NMeth}. 

To increase the probability of BH finding a stable global optimum, we modify the \texttt{RandomPerturbation} step of Algorithm \ref{alg:cap} as follows:
\begin{enumerate}
    \item If any parameters in $\Theta$ are restricted to the positive domain, \texttt{RandomPerturbation} is repeated until all positivity constraints in $\Theta'$ are satisfied.
    \item Instead of utilizing a constant stepsize $\alpha$, a random stepsize is drawn from $[0,\alpha]$ each time \texttt{RandomPerturbation} is called.
\end{enumerate}
Our first modification to BH restricts the BH proposal to the appropriate domain of $\chi^2_{\mathrm{aug.}}(\Theta)$. Our second modification empirically increases the rate at which BH finds a stable optimum. 

At the \texttt{LocalOptimization} step of Algorithm \ref{alg:cap}, we optimize $\Theta'$ using the trust region reflective algorithm implemented in \texttt{SciPy} \cite{branch1999subspace, 2020SciPy-NMeth}. Though not reported in this work, we find that the trust region reflective algorithm is very robust and outperforms many popular local optimization algorithms, such as L-BFGS-B, Levenberg-Marquardt, and conjugate gradient \cite{byrd1995limited, levenberg1944method, marquardt1963algorithm, hestenes1952methods}. Gradients of $\chi^2_{\mathrm{aug.}}$ with respect to $\Theta$ are calculated using the automatic differentiation tools provided by the \texttt{GVar} library \cite{gvar}. 

\begin{figure}
    \centering
    \includegraphics[width=\columnwidth]{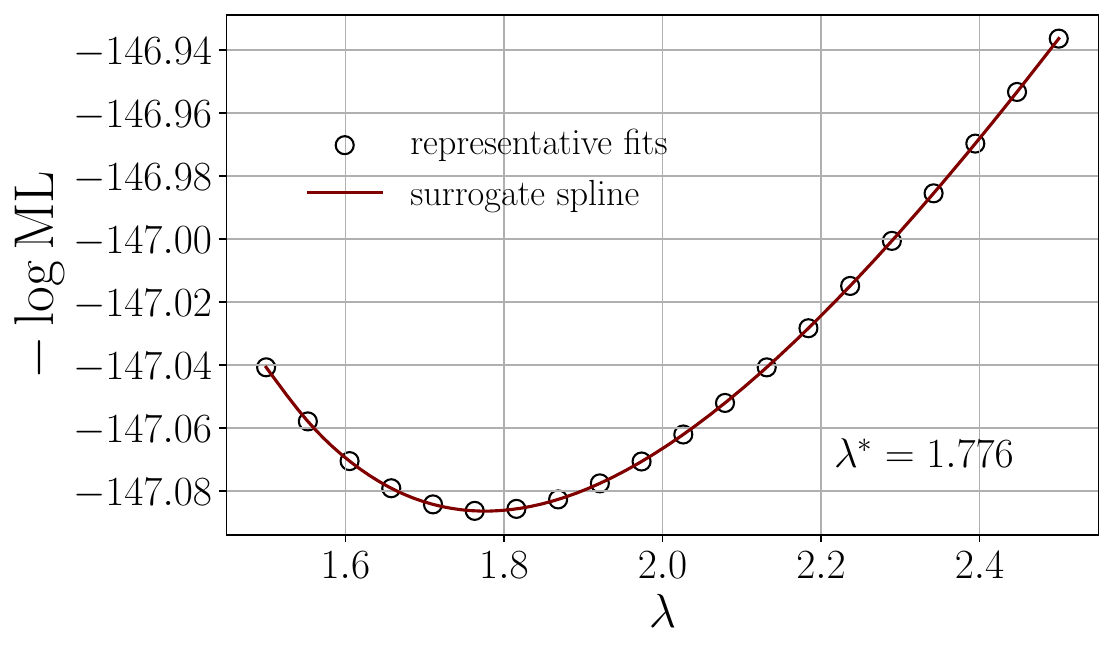}
    \caption{Example of an interpolation over the marginal likelihood in Eqn. \ref{eqn:marginal_likelihood} with a cubic spline (red line). Each black circle represents the marginal likelihood calculated from a curve collapse fit of the 2-state Potts model Binder cumulant at a particular value of $\lambda$ (Eqn. \ref{eqn:ridge_prior}). The minimum of the surrogate spline $\lambda^{*}$ is the value for $\lambda$ suggested by the empirical Bayes procedure.}
    \label{fig:surrogate_spline}
\end{figure}

The basin hopping algorithm is terminated after $\Theta_{\mathrm{best}}$ in Algorithm \ref{alg:cap} has remained the same for a pre-set number of iterations. We typically start the algorithm off with a large stepsize $\alpha$, then we tune $\alpha$ to smaller values as $\chi^2_{\mathrm{aug.}}(\Theta)$ decreases until the algorithm stabilizes its sampling of $\chi^2_{\mathrm{aug.}}(\Theta)$. We wish to explore a better procedure for automatically tuning $\alpha$ throughout the $\chi^2_{\mathrm{aug.}}(\Theta)$ optimization procedure in the future. Additionally, there may be better step-taking procedures that can exploit the structure and symmetries of artificial neural networks to more efficiently explore the $\chi^2_{\mathrm{aug.}}$ landscape. 

\section{Surrogate-based empirical Bayes}\label{sec:empirical_bayes}

We calculate $\lambda$ in Eqns. \ref{eqn:ridge_prior} and \ref{eqn:poly_ridge_prior} using the empirical Bayes method discussed in Ref. \cite{Lepage:2001ym}. Empirical Bayes picks out the value of $\lambda$ that extremizes the marginal likelihood; as such, the model defined by the empirical Bayes procedure is an approximation to a full hierarchical Bayesian model \cite{murphy2023probabilistic}. We approximate the marginal likelihood using a Laplace approximation of the posterior distribution, yielding
\begin{equation}\label{eqn:marginal_likelihood}
    -2\log \mathrm{ML} \approx \chi^2_{\mathrm{aug.}}(\Theta^{*})+\log \frac{(2\pi)^{\mathrm{d.o.f.}}\det \Sigma_{O}\det \Sigma_{C}}{\mathrm{det}\Sigma_{\Theta^{*}}}.
\end{equation}
As $\log\mathrm{ML}$ is a function of $\lambda$, each value of $\log\mathrm{ML}$ is calculated by performing a fit. Therefore, it is computationally advantageous to reduce the number of fits needed to be performed to optimize $\log\mathrm{ML}$. We do so via the following three-step procedure:
\begin{enumerate}
    \item Perform $M$ fits in the range $[\lambda_{\mathrm{min.}},  \lambda_{\mathrm{max.}}]$ and estimate $\log\mathrm{ML}$ using Eqn. \ref{eqn:marginal_likelihood} for each fit. 
    \item Interpolate over the estimate of $\log\mathrm{ML}$ in $\lambda$ for each fit using a cubic spline. We use either the monotonic or smooth spline algorithm provided by \texttt{GVar} \cite{steffen1990simple, gvar}.
    \item Optimize $\log\mathrm{ML}$ by utilizing the cubic spline as a surrogate for $\log\mathrm{ML}$.
\end{enumerate}
The accuracy of $\lambda$ scales with the number of fits $M$ entering the knots of the cubic spline. Very few fits are typically needed to obtain a reasonable estimate for the $\lambda=\lambda^{*}$ that optimizes $\log\mathrm{ML}$. In Fig. \ref{fig:surrogate_spline}, we show an example of this procedure for our curve collapse analysis of the Binder cumulant in the $2$-state Potts model discussed in Sec. \ref{sec:curve_collapse}. Each open circle represents the $\log\mathrm{ML}$ calculated from a fit at a particular value of $\lambda$. The red line is an interpolation of $\log\mathrm{ML}$ in $\lambda$ using a cubic spline. The value $\lambda^{*}=1.776$ that optimizes $\log\mathrm{ML}$ is calculated by minimizing the value of $-\log\mathrm{ML}$ that we estimate from the cubic spline. Though our surrogate-based empirical Bayes procedure can be implemented in an embarrassingly parallel fashion, it is numerically more stable to calculate the marginal likelihood by starting at a small value of $\lambda$, then calculate subsequent marginal likelihoods at larger $\lambda$ by sequentially initializing each fit with the parameters calculated from the previous fit.

It is worth noting that this procedure could be improved by nesting it into an iterative bisection-based optimization algorithm, whereby $\lambda_{\mathrm{min.}},\lambda_{\mathrm{max.}}$ converge to $\lambda^{*}$ using the derivative of the spline-based surrogate of $\log\mathrm{ML}$ as the number of iterations increases. The present procedure may also be extended to multiple dimensions; however, it is probably more efficient to utilize a Bayesian optimization algorithm for multi-dimensional optimization of $\log\mathrm{ML}$.

\section{Interpolation}\label{sec:interpolation}

\begin{figure}
    \centering
    \includegraphics[width=\columnwidth]{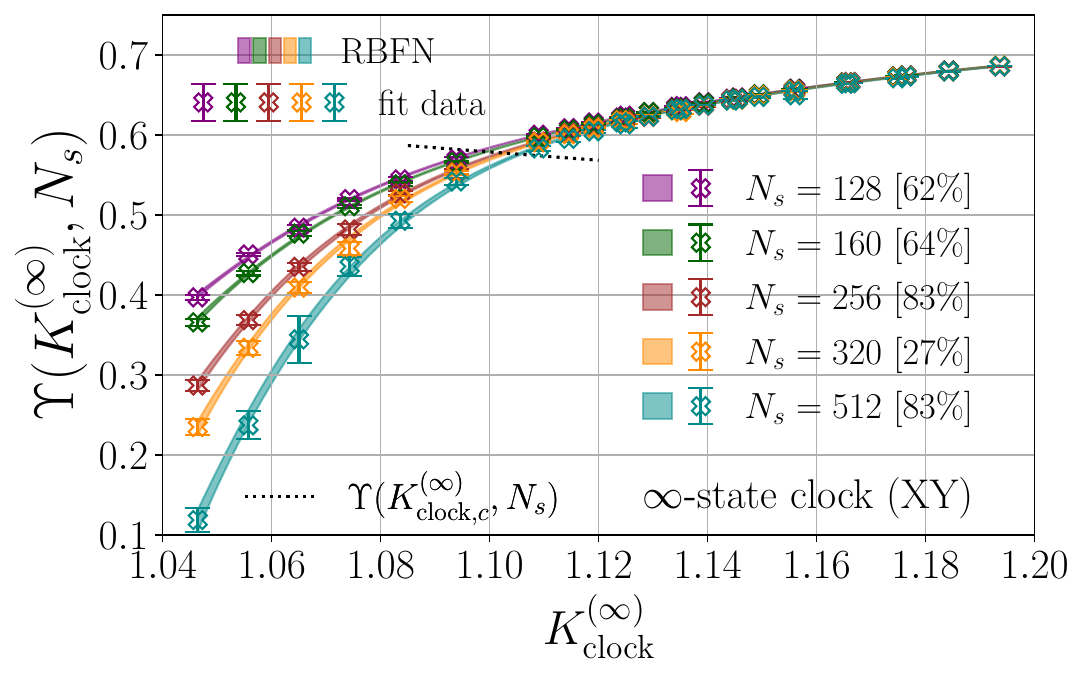}
    \caption{RBFN-based interpolation of the helicity modulus $\Upsilon(K_{\mathrm{clock}}^{(\infty)},N_{s})$ for the $\infty$-state clock (XY) model at fixed $N_{s}$. Data included in fit is shown as an errorbar with an open ``$\times$'' marker. RBFN-based interpolation is shown as a colored band. Interpolation performed on $N_{s}=128$ (purple), $160$ (dark green), $256$ (red), 320 (yellow), and $512$ (cyan). The RBFN-based fits are shown as a colored bands, with the width of the band indicating the error. The color of each band indicates the $N_{s}$ at which the fit was performed. The helicity modulus at $K_{\mathrm{clock},c}^{(\infty)}(N_{s})$ given by Eqn. \ref{eqn:jmp} is indicated by a dotted black line. The RBFN has 2 nodes in its hidden layer.}
    \label{fig:xy_helicity_modulus}
\end{figure}

\begin{figure}
    \centering
    \includegraphics[width=\columnwidth]{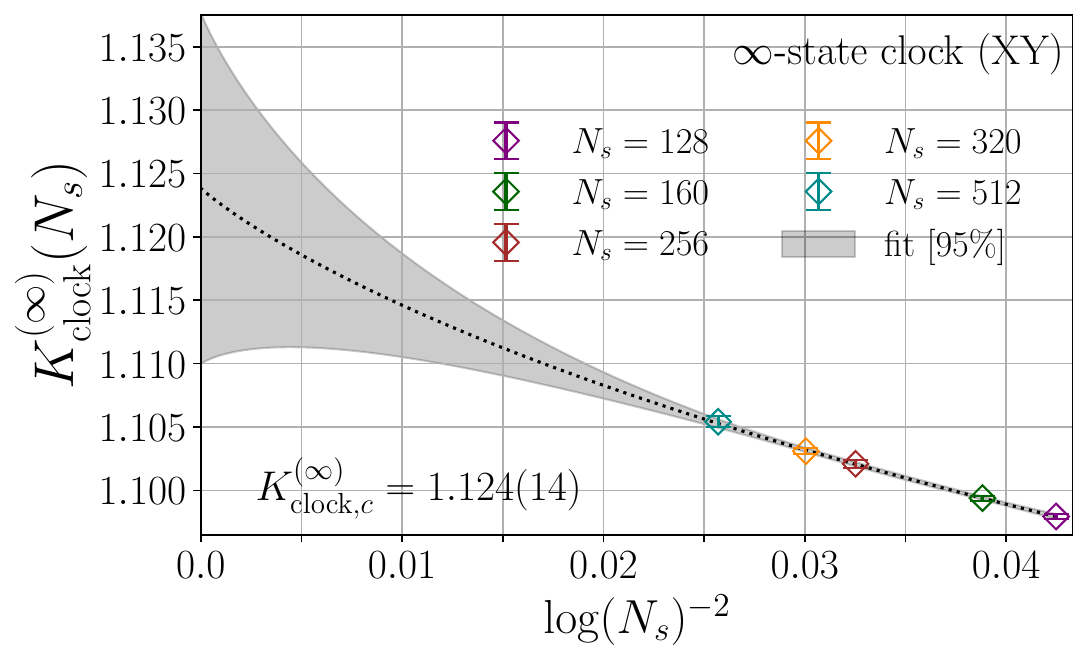}
    \caption{Extrapolation of the pseudocritical temperature $K_{\mathrm{clock},c}^{(\infty)}(N_{s})$ calculated from the intersection of our RBFN-based interpolation (colored bands in Fig. \ref{fig:xy_helicity_modulus}) with the universal jump condition (dotted line in Fig. \ref{fig:xy_helicity_modulus}) to $1/N_{s} \rightarrow 0$ using Eqn. \ref{eqn:bkt_extrap}. The pseudocritical temperatures are indicated by multi-colored errorbars with open diamond markers ``$\diamond$'' and utilize the same color scheme as Fig. \ref{fig:xy_helicity_modulus} for different $N_{s}$ (see caption). Result of fit to Eqn. \ref{eqn:bkt_extrap} is shown as a grey band and the central value of the fit prediction is shown as a dotted black line.}
    \label{fig:xy_tc_extrapolation}
\end{figure}

In this appendix, we demonstrate the use of RBFNs for direct interpolation. We calculate $K_{\mathrm{clock},c}^{(\infty)}$ for the $\infty$-state clock (XY) model using the helicity modulus
\begin{equation}
    \Upsilon\big(K_{\mathrm{clock}}^{(p)}, N_{s}\big)=\frac{1}{2}\sum_{\mu=1}^{2}\big\langle e_{\mu} - N_{s}^{2}K_{\mathrm{clock}}^{(p)}s_{\mu}^2 \big\rangle,
\end{equation}
with
\begin{gather}
   e_{\mu}=\frac{1}{N_{s}^{2}}\sum_{\langle ij \rangle_{\mu}}\cos(\theta_i-\theta_j), \\
   s_{\mu}=\frac{1}{N_{s}^{2}}\sum_{\langle ij \rangle_{\mu}}\sin(\theta_i-\theta_j),
\end{gather}
where $\langle ij \rangle_{\mu}$ denotes a sum of lattice sites $i$ along the $\mu$-direction and their nearest-neighbors $j$ \cite{van1981helicity, komura2012large, nguyen2021superfluid, tuan2022binder}. We fit the data for $\Upsilon\big(K_{\mathrm{clock}}^{(\infty)}, N_{s}\big)$ at fixed $N_{s}$   with an RBFN  that possess 2 nodes in its hidden layer. The fits to $\Upsilon\big(K_{\mathrm{clock}}^{(\infty)}, N_{s}\big)$ yield  p-values in the $27\%-83\%$ range, as shown in Fig. \ref{fig:xy_helicity_modulus}. At $K_{\mathrm{clock},c}^{(\infty)}$, there is a universal jump condition
\begin{equation}\label{eqn:jmp}   \Upsilon\big(K_{\mathrm{clock},c}^{(p)},N_{s}\big)=2f_{r}/\pi K_{\mathrm{clock},c}^{(p)}(N_{s}),
\end{equation}
where $f_{r}=1-16\pi\exp(-4\pi)$ \cite{nguyen2021superfluid}. The universal jump condition is shown as a dotted black line in Fig. \ref{fig:xy_helicity_modulus}. By calculating where our RBFN-based interpolation of $\Upsilon\big(K_{\mathrm{clock}}^{(\infty)}, N_{s}\big)$ intersects the universal jump condition, we can calculate the pseudocritical temperature $K_{\mathrm{clock},c}^{(\infty)}(N_{s})$. From $K_{\mathrm{clock},c}^{(\infty)}(N_{s})$ at multiple $N_{s}$, we extrapolate to $1/N_{s} \rightarrow 0$ using the ansatz 
\begin{equation}\label{eqn:bkt_extrap}
    K_{c}(N_{s})=K_{c}+\zeta^{-1/\nu}\log\big(\kappa N_{s}\big)^{-1/\nu},
\end{equation}
where $\kappa$ is a free parameter and $\nu$ set to its exact value of $\nu=1/2$ \cite{Kosterlitz:1974sm}, as shown in Fig. \ref{fig:xy_tc_extrapolation}. Our extrapolation yields a prediction for $K_{\mathrm{clock},c}^{(\infty)}$ that is consistent with the estimates of Refs. \cite{hasenbusch2005two, komura2012large, Sale:2021xsq, nguyen2021superfluid} at the $1\sigma$ level, though with considerable statistical uncertainty due to the logarithmic scaling of $K_{\mathrm{clock},c}^{(\infty)}(N_{s})$ with $N_{s}$. 

\end{document}